\PassOptionsToPackage{numbers,sort&compress}{natbib}
\documentclass[final,5p,times,twocolumn]{elsarticle}

\usepackage{amssymb}
\usepackage{amsmath}
\usepackage{graphicx}
\usepackage{bm}        

\usepackage{slashed}
\setcitestyle{numbers,sort&compress}
\usepackage{hyperref}

\journal{Physics Letter B}

\begin{document}

\begin{frontmatter}



\title{Probing the CP-Violation effects in the $h\tau\tau$ coupling at the LHC}


\author[addr1,addr2,addr3]{Xin Chen}
\ead{xin.chen@cern.ch}

\author[addr4]{Yongcheng Wu}
\ead{ycwu@physics.carleton.ca}

\address[addr1]{Department of Physics, Tsinghua University, Beijing 100084, China}
\address[addr2]{Collaborative Innovation Center of Quantum Matter, Beijing 100084, China}
\address[addr3]{Center for High Energy Physics, Peking University, Beijing 100084, China}
\address[addr4]{Ottawa-Carleton Institute for Physics, Carleton University, \\
                          1125 Colonel By Drive, Ottawa, Ontario K1S 5B6, Canada}

\begin{abstract}
A new method used to calculate the neutrino for all major tau hadronic decay event by event at the LHC is presented. It is possible because nowadays better detector description is available. With the neutrino fully reconstructed, matrix element for each event can be calculated, the mass of the Higgs particle can also be calculated event by event with high precision.
Based on these, the prospect of measuring the Higgs CP mixing angle with $h\to\tau\tau$ decays at the LHC is analyzed. 
It is predicted that, with a detailed detector simulation, with 3 ab$^{-1}$ of data at $\sqrt{s}=13$ TeV, a significant improvement of the measurement of the CP mixing angle to a precision of $5.2^\circ$ can be achieved at the LHC, which outperforms the sensitivity from lepton EDM searches up to date in the $h\tau\tau$ coupling.
\end{abstract}


\end{frontmatter}

\section{Introduction}
To account for the large asymmetry between the matter and anti-matter in our Universe, enough CP violation effects should be presented in the theory. However, in the Standard Model, the CP phase in the CKM matrix is not sufficient for this purpose. New physics 
is therefore needed to introduce more CP violation sources. Possible candidates are Supersymmetry, Left-Right Symmetric model, etc.

On the other hand, the new discovered Higgs boson also opens a window towards the new physics. The precision measurement of Higgs properties will be one of the most important targets of the LHC in the next running periods. Among them, the CP property is an important topic. The pure CP eigenstate assumption has already been investigated at the LHC experiments~\cite{ATLAS-hzzCP,CMS-hzzCP1,CMS-hzzCP2} in the diboson decays, and the pure CP-odd situation is excluded better that $99.9\%$ CL. The $h\to ZZ^*\to 4l$ is the golden channel for this measurement, subject to the scale suppression due to dim-6 operators, whereas for the Yukawa coupling, $h\to\tau\tau$ is the best channel we could use and is widely investigated in the literatures~\cite{Han:2016bvf,cepc_cp,Hagiwara:2016zqz,DellAquila:1988bko,He:1993fd,Hayreter:2016kyv,Hayreter:2016aex,Bower:2002zx,Desch:2003mw,Harnik:2013aja,Berge:2008wi,Berge:2008dr,Berge:2011ij,Berge:2013jra,Berge:2015nua,Dolan:2014upa,Askew:2015mda,Kuhn,imp1,imp2}. However, the missing neutrino from the decay of each tau makes it difficult to achieve a better precision on measuring the CP mixing angle ($\phi$) of the $h\tau\tau$ interaction which we assume to have the following effective form: 
\begin{equation}
\label{equ:Lhtt}
\mathcal{L} = -y_\tau\overline{\tau}(\cos\phi + i\gamma_5\sin\phi)\tau h,
\end{equation}


In this work, utilizing the mass constraints and impact parameters, the new method in \cite{cepc_cp} to reconstruct neutrinos with impact parameters and resolutions taken into account is used, which is combined with the matrix element in the LHC setting for the first time.
With the momentum of the neutrino from the tau decay fully reconstructed within certain accuracy, an observable based on the matrix element to retrieve the CP information in the $h\tau\tau$ interaction can be calculated, which can achieve higher precision as it contains more information in the final states integrated in the matrix element.
Under this framework, all major tau hadronic decay modes are included and combined in this work simultaneously, which gives the important prediction on the best we can do with the Higgs CP in its fermionic coupling at the LHC. 
Similar to \cite{ee_to_WW}, the energy-angle correlations of the final tau lepton decay products are used to extract the CP information.
A detailed simulation study shows that a significant improvement of the measurement of the CP mixing angle can be achieved. The method is illustrated using the measurement of CP-violation effects in $h\tau\tau$ interaction as an example, extension to other situation containing tau decay should be straightforward and will further improve the precision.


In this work, the track trajectory is assumed to follow a helix, as opposed to an idealized straight line assumed in \cite{Kuhn,Berge:2015nua}. 
An explicit method for neutrino reconstructions started in \cite{imp2}, which used the impact parameters and track momentum to define a track plane, on which subsequent calculations are based. On the other hand, in this work, the impact parameters $d_0$ and $z_0$ are used as auxiliary measurements which are related to the tau flight direction, and there is no decay plane explicitly formed as done in \cite{Berge:2015nua,imp2}, which suffers from impact parameters resolutions. For the multi-prong tau decay, no common vertex is attempted for the multi-prongs in our method. This vertex suffers from large uncertainty because not large number of tracks are available for a precise vertex determination. In these different aspects, the method in \cite{cepc_cp} is considered new, and is first applied to hadron collider settings in this work with the helix shape simulated for tracks by DELPHES 3.4.0~\cite{Delphes}.

To probe the Higgs CP angle, a sample of events with high signal purity are needed. Therefore, the analysis is carried out in the Vector-Boson-Fusion (VBF) production mode of Higgs, as opposed to the gluon-gluon-Fusion mode used in \cite{Dolan:2014upa} and other places, whose signal to background ratio is much worse than VBF \cite{CMS_htt,ATLAS_htt}. This choice is also in line with \cite{Han:2016bvf}.

\section{Reconstruction of the neutrino at the LHC}

We will first describe the simulation detail for the measurement of CP-violation effects in $h\tau\tau$ interaction and then illustrate our method in this situation in detail.

\subsection{Monte-Carlo Simulation of the signal and background processes}

The VBF channel will be used for the Higgs production, as already been investigated in~\cite{Englert:2015dlp,Han:2016bvf}, it is the most promising production channel for CP analysis. Although the gluon-gluon fusion is the dominant production channel for the Higgs at the LHC, it has larger background and lower signal purity, which will impact a lot the measurement of the CP property~\footnote{In~\cite{Hagiwara:2016zqz}, the authors use the ggF channel to check the precision of the measurement. In our opinion, it can be improved by using the VBF channel.}. Only major hadronic decay modes for two taus from Higgs decay are used, while the leptonic decay mode is excluded as it contains extra missing neutrino, which will induce a new unknown parameter that is hard to construct, and the hadronic modes already have sufficient statistics~\cite{CMS_htt,ATLAS_htt}. 
The main backgrounds for this signal come from the $Z$ production associated with additional jets. The processes we consider are listed in following:

\begin{itemize}
\item Signal: $p\ p\to h\ j\ j$, $h\to \tau\ \tau$.
\item Background:  $p\ p\to Z + 0,1,2,3 j$, $Z \to \tau\ \tau$ (QCD $Z$+jets).
\item Tau Decay Modes Used: \begin{itemize}
\item $\tau^\pm\to\pi^\pm\nu$,
\item $\tau^\pm\to\rho^\pm\nu\to\pi^\pm\pi^0\nu$,
\item $\tau^\pm\to a^\pm\nu\to\pi^\pm\pi^\pm\pi^\mp\nu$,
\end{itemize} 
\end{itemize} 
and all six combinations of the tau decay modes are used in our analysis. The other backgrounds, mainly dominated by the QCD, is also important~\cite{CMS_htt,ATLAS_htt}. However, QCD fake background is beyond the scope of this work. 
We will just assume the same cross section after all selection cuts as the QCD $Z$+jets background for simplicity, which is roughly consistent with the current results in \cite{CMS_htt,ATLAS_htt}.

The VBF signal is generated with Powheg~\cite{Powheg} at NLO accuracy in QCD 
and PDF set NNPDF30NLO~\cite{Ball:2014uwa}, 
and interfaced to Pythia8~\cite{Pythia8} for resonance decays, parton shower and hadronization. The QCD and EW $Z$+jets background
is generated at LO with MadGraph5~\cite{MG5} 
and PDF set NNPDF23LO~\cite{Ball:2013hta}, 
with up to three extra partons. Samples with different parton multiplicities are merged according to the CKKW-L method~\cite{CKKW-L}, and showered by Pythia8. A k-factor of 1.23 is applied to the QCD $Z$+jets cross section to match the NNLO prediction~\cite{Z_xsec}. The spin correlation between two taus is retained during the decays by Pythia. The events are afterwards passed through DELPHES simulating the detector response of the ATLAS detector at HL-LHC \cite{HL_LHC}. 

The tracking range is defined to be consistent with the current ATLAS detector ($|\eta|<2.5$)\footnote{Although in HL-LHC, it is expected that the tracking range will be extended to $|\eta|<4.0$ with silicon trackers for ATLAS, in this work to be conservative, the tracking is still limited to $|\eta|<2.5$ consistent with the current detector layout.}. The track and calorimeter resolutions, the track finding and lepton identification efficiencies are taken from the default ATLAS parameter cards in DELPHES 3.4.0.
Charged tracks have an efficiency of $92\%$ ($87\%$) in the $|\eta|\leq 1.5$ ($1.5<|\eta|<2.5$) region. Charged hadron momentum resolutions are $0.9\%\oplus 1.8\times 10^{-4} p_\text{T}$ ($1.8\%\oplus 2.4\times 10^{-4} p_\text{T}$) for tracks in $|\eta|\leq 1.0$ ($1.0<|\eta|<2.5$), where $p_\text{T}$ is in GeV \cite{tracking}.
The jets are formed 
based on the Anti-$k_{t}$ algorithm~\cite{antikt} with a cone parameter of 0.4
\footnote{For jet energy and missing transverse energy calculations, the EM calorimeter resolutions are parametrized as $10.1\%\sqrt{E}\oplus 0.17\%E$ and $28.5\%\sqrt{E}\oplus 3.50\%E$ for $|\eta|\leq 3.2$ and $3.2<|\eta|<4.9$, respectively. The hadronic calorimeter resolutions are $1.59\oplus 52.05\%\sqrt{E}\oplus 3.02\%E$, $70.6\%\sqrt{E}\oplus 5.00\%E$ and $100.0\%\sqrt{E}\oplus 9.42\%E$ for $|\eta|\leq 1.7$, $1.7<|\eta|\leq 3.2$ and $3.2<|\eta|<4.9$, respectively. The energy $E$ is all in GeV.}
. The hadronic tau tagging is performed on these jets with an efficiency of $75\%$ ($60\%$) and fake rate of $4\%$ ($0.4\%$) for 1-prong (3-prong) real and fake tau objects, respectively \cite{TauCP_HL}. Identification of different tau decay modes is essential, the development of tau substructure algorithms \cite{tausub1,tausub2} improved the tau energy resolution by a factor of two with respect to the previous calorimeter-based algorithms, and neutral pion's energy can be resolved to $16\%$. In this work, we assume that different tau decay modes can be classified without crosstalk, and the neutral pion energy can be resolved with $15\%$ uncertainty for the HL-LHC period.

The impact parameters of the tracks are used to constrain the neutrino momenta from tau decays, as used in~\cite{cepc_cp}. A simple resolution of the form $a\oplus b/(p_\textrm{T}\sin^{1/2}\theta)$, where $p_\textrm{T}$ is in GeV and $\theta$ is the polar angle of the track, is applied. The parameters of $a=8$ (9) $\mu$m and $b=70$ (80) $\mu$m are set for tracks in $|\eta|\leq 1.0$ ($1.0<|\eta|<2.5$) for the resolutions of $d_0$. For $z_0$, $a=10$ (20) $\mu$m and $b=90$ (200) $\mu$m are set for tracks in $|\eta|\leq 1.0$ ($1.0<|\eta|<2.5$) \cite{tracking}.
It is further assumed that the resolution of the primary vertex can be precisely resolved by the multiple tracks from the VBF jets and underlying event, and the additional ``smearing" due to interaction point uncertainty is not considered.

%

\subsection{Reconstruction of the missing neutrino}

\begin{figure}[!tb]
\centering
\includegraphics[width=0.5\textwidth]{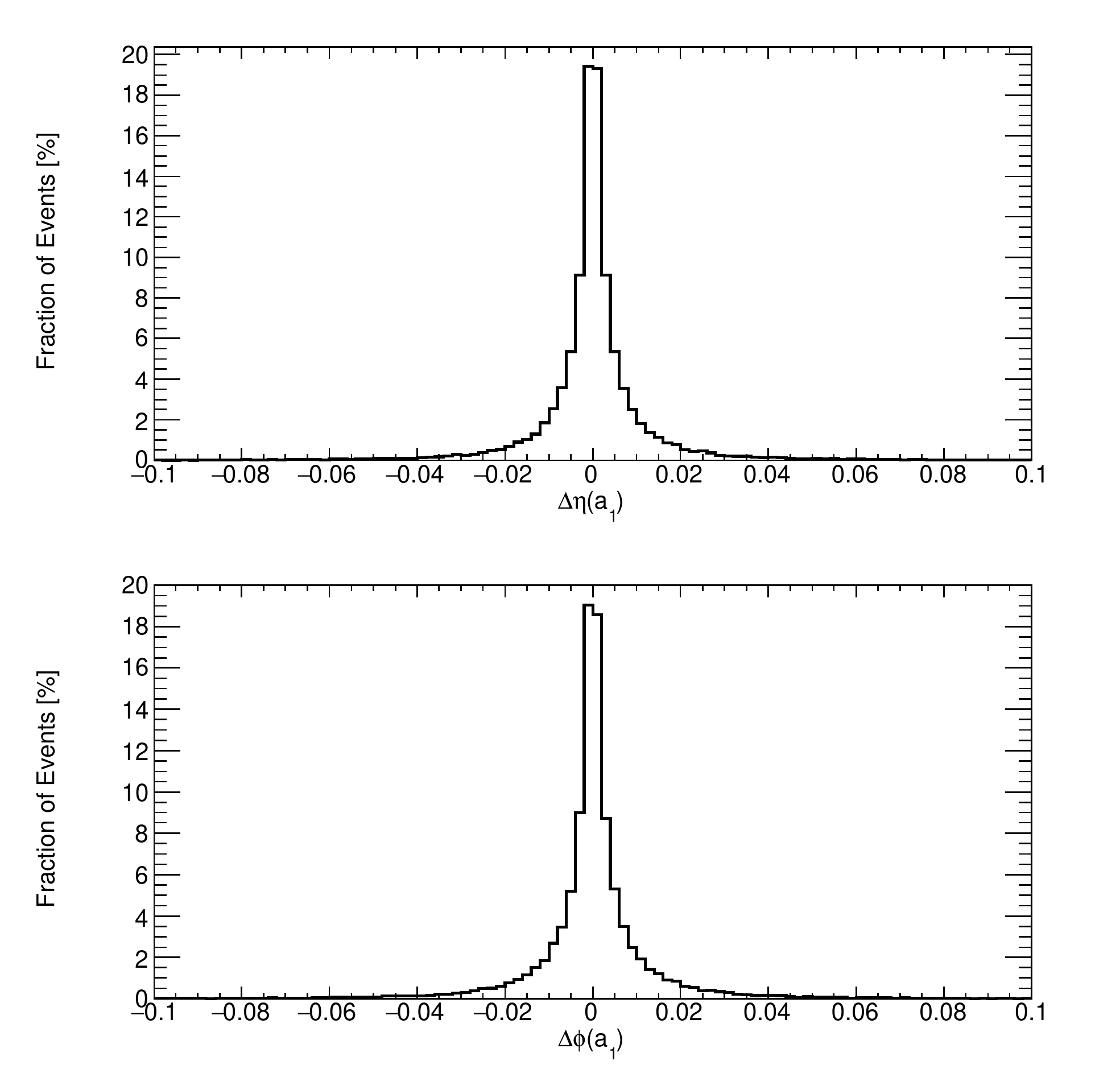}
\caption{The difference in $\eta$ and $\phi$ between the fitted and true values for the taus which decay via $a_1 \nu$.}
\label{fig:a1}
\end{figure}

In each event, the two missing neutrinos contain six free parameters, while in this case the mass peak, missing energy and impact parameter measurement can provide at least seven constraints which are sufficient to estimate the six free parameters with some uncertainties. In this work, minimum chi-square fitting is used, and each constraint will contribute one term to the total $\chi^2$ which has the form of

%
\begin{eqnarray}
\begin{large}
\begin{array}{ll}
  \chi^2 = &  \left( \frac{m_{\tau\tau}^{\footnotesize{\textrm{fit}}} - m_h}{\sigma_{h}} \right)^2 + \left(\frac{m_{\tau 1}^{\textrm{fit}}-m_\tau}{\sigma_{\tau}}\right)^2 + \left(\frac{m_{\tau 2}^{\textrm{fit}}-m_\tau}{\sigma_{\tau}}\right)^2 + \\
  &  \left(\frac{\slashed{E}_{x}^{\textrm{fit}} - \slashed{E}_{x}}{\sigma_{\textrm{mis}}} \right)^2 + \left(\frac{\slashed{E}_{y}^{\textrm{fit}} - \slashed{E}_{y}}{\sigma_{\textrm{mis}}} \right)^2  + \chi^2_{\textrm{Imp}}
\end{array}
\end{large}
\label{eq:eq4}
\end{eqnarray}
The first three terms are provided by the mass constraints from the Higgs mass and the two tau mass, where $m_h=125$ GeV, $m_\tau=1.777$ GeV, $\sigma_{h}=10$ GeV, $\sigma_{\tau}=0.1$ (0.2) GeV for taus decaying to $a_1\nu$ or $\pi\nu$ ($\rho\nu$). The 4th and 5th terms come from the missing energy measurement, where $\slashed{E}_{x,y}$ is the missing transverse energy, $\sigma_{\textrm{mis}}=0.67\sqrt{\Sigma E_{\textrm{T}}/\textrm{GeV}}$ is its resolution. The last term represents the constraint from the impact parameters of the tracks, which has the form $\chi^2_{\textrm{Imp}}  =  \sum_i \chi_{\textrm{Imp},i}^2$, with
\begin{equation}
\chi^2_\text{Imp,i}  =  \left(\frac{d_0^{\text{fit}}-d_0}{\sigma_{d_0}}\right)^2 + \left(\frac{z_0^{\text{fit}}-z_0}{\sigma_{z_0}}\right)^2 ,\\
\label{eq:eq6}
\end{equation}
and the details can be found in \cite{cepc_cp}.  The variables with a superscript ``fit" are fitted variables, which incorporate the neutrino 4-momenta that need to be determined by minimizing Eq. \ref{eq:eq4}. The 3-prong mode is special since each track contributes a constraint as Eq. \ref{eq:eq6}. Minimizing their sum can directly give the flight direction of the tau lepton\footnote{The impact parameters of each track gives one constraint, and three constraints are sufficient to determine the tau flight direction which has two unknowns.}. Figure \ref{fig:a1} shows the difference between the fitted and true tau flight directions (pseudorapidity $\eta$ in top panel and azimuthal angle $\phi$ in bottom panel) before its decay. For the 3-prong mode, tau flight direction ($\eta_\tau,\phi_\tau$) is first obtained by minimizing $\sum_{i=1}^{3}\chi^2_{\text{Imp},i}$, and a new term of the form
\begin{equation}
\chi^2_\text{Imp} = \left(\frac{\eta_\tau^{\text{fit}} - \eta_\tau}{\sigma_\eta}\right)^2 + \left(\frac{\phi_\tau^{\text{fit}} - \phi_\tau}{\sigma_\phi}\right)^2 ,
\label{eq:eq7}
\end{equation}
where $\sigma_\eta=\sigma_\phi=0.007$ based on the results in Fig. \ref{fig:a1}, replaces the sum of $\chi^2_{\text{Imp},i}$ in the last term of Eq. \ref{eq:eq4} in the per-event minimization.

For the final states with an intermediate $\rho$ meson, extra terms,
\begin{equation}
\left(\frac{m_\rho - 0.775}{0.2}\right)^2 + \left(\frac{f_{\pi^0} - 1}{0.15}\right)^2 ,
\label{eq:eq8}
\end{equation}
where $f_{\pi^0}$ is the energy scale factor applied on the $\pi^0$ 4-momentum, are added to Eq.~\ref{eq:eq4}. This term reflects the additional uncertainty due to the neutral cluster resolution.

In the per-event minimization of Eq.~\ref{eq:eq4}, the $\eta_\nu^{\textrm{fit}}$ and $\phi_\nu^{\textrm{fit}}$ of one neutrino are first scanned over, from which the magnitudes of the neutrinos' momenta and the direction of the other neutrino can be obtained via the tau mass and $\slashed{E}_{x,y}$ constraints in Eq.~\ref{eq:eq4}. Conversely, the scan is repeated starting from the parameters of the other neutrino. After a coarse global minimum is found by the scan, a fit using MINUIT~\cite{MINUIT} is performed around this minimum point for a better estimation. 

\begin{figure}[!tb]
\centering
\includegraphics[width=0.50\textwidth]{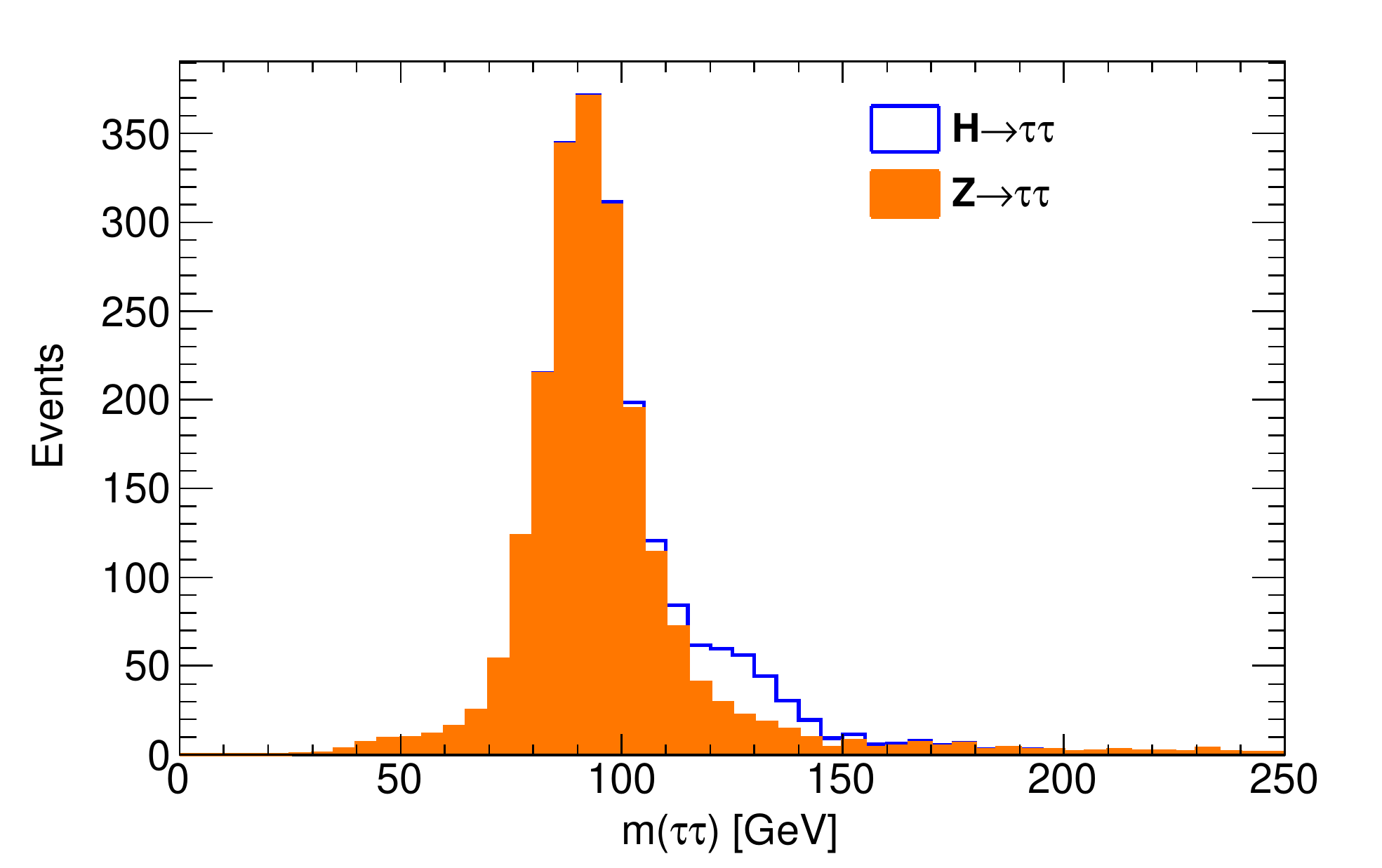}
\caption{The expected unconstrained $m_{\tau\tau}$ distribution with 300 fb$^{-1}$ after the VBF cuts with all channels combined.}
\label{fig:mass}
\end{figure}

\begin{figure}[!tb]
\centering
\includegraphics[width=0.50\textwidth]{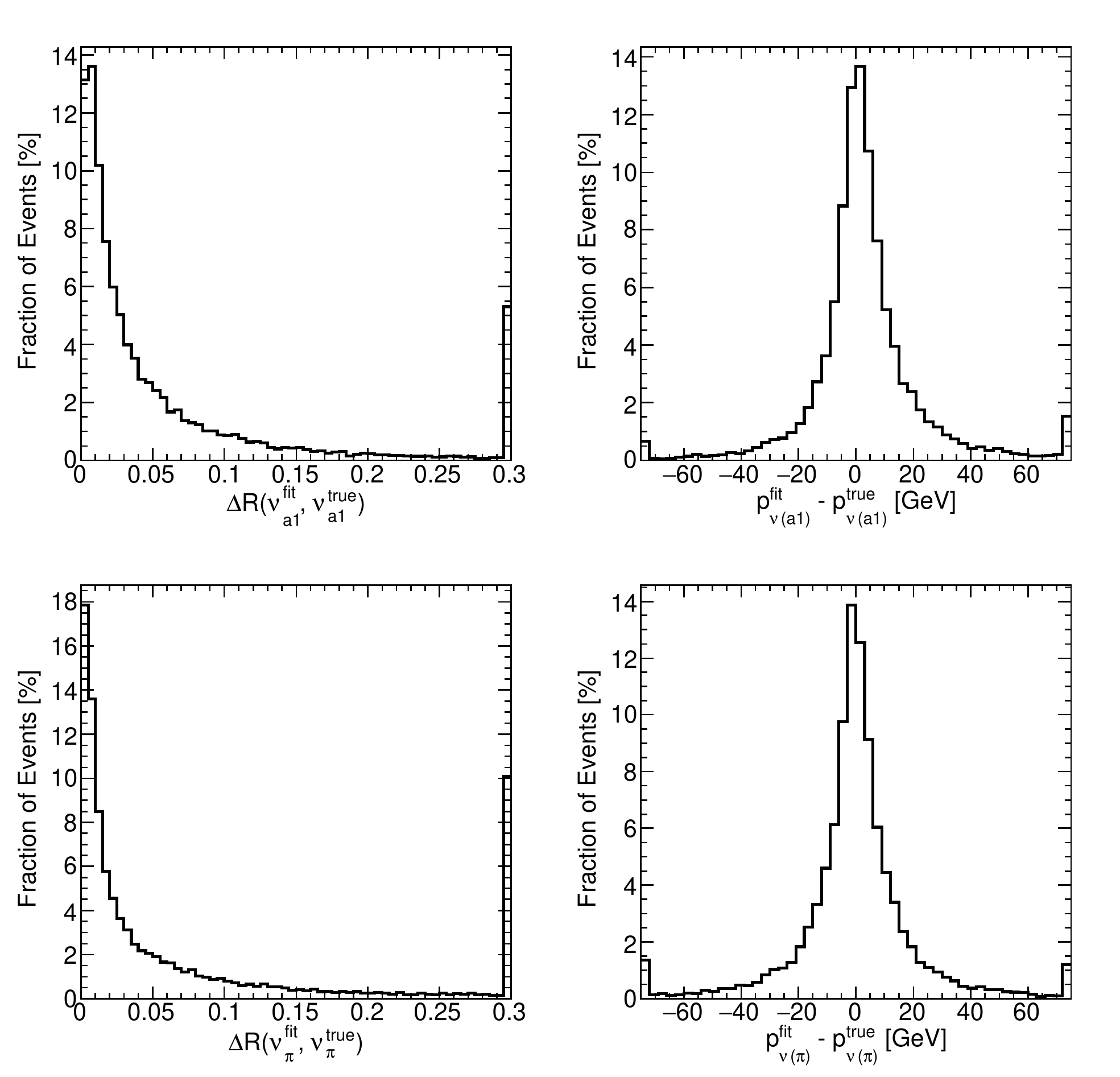}
\caption{The $\Delta R$ and difference between fitted and true neutrino momenta from the hadronic tau decays in the $a_1\nu$+$\pi\nu$ channel. The rightmost bins in all plots and the leftmost bins in the right-column plots indicate the overflows. }
\label{fig:reco}
\end{figure}

With the Higgs mass constraint term in Eq.~\ref{eq:eq4}, the background ditau mass is also biased to the nominal Higgs mass at 125 GeV.  To select a pure signal sample and extract the CP information, a two-step procedure is adopted.
\begin{enumerate}
\item In the first step, the fit is done without the Higgs mass constraint term in Eq.~\ref{eq:eq4}. The distribution of unconstrained $m_{\tau\tau}$ is shown in Fig.~\ref{fig:mass} and will be used to select events.
\item In the second step, after the selection cuts in Sec. \ref{sec:selection}, the Higgs mass constraint term is put back in Eq.~\ref{eq:eq4}, and the fit is done to extract the neutrinos' information with higher precision. The $\Delta R$ and momentum ratio between the fitted and true neutrinos in the $a_1 + \pi$ channel are shown in Fig.~\ref{fig:reco}. It is seen that good neutrino direction can be obtained, and the resolution for the neutrino momentum magnitude is about 8 GeV. 
Other channels have similar precisions and are not shown here.
\end{enumerate}

\subsection{Event selections}
\label{sec:selection}

Following cuts \cite{VBF_pheno,CMS_htt,ATLAS_htt} are used to select reconstructed events to achieve better signal to background ratio:  
\begin{description}
\item[Tau cuts] The tau candidate should have one or three tracks with a unit charge. The leading track has $p_{\textrm{T}}>5$ GeV. For the 3-prong tau, $p_{\textrm{T}}>2$ GeV on the other tracks. The two taus have opposite charge, and are within $|\eta|<2.5$. To take into account the trigger, $p_{\textrm{T}}>40,30$ GeV are required on the two taus. They should also have $|\Delta\phi|<2.9$ to avoid the back-to-back topology.
\item[VBF Cut] $p_{\textrm{T}}^{j_1}>50\textrm{ GeV},\ p_{\textrm{T}}^{j_2}>40 \textrm{ GeV}$, \\ $|\Delta\eta_{jj}|>3.8,\ m_{jj}>500\textrm{ GeV},\ \eta_{j_1}\times\eta_{j_2}<0$,
\item[Tau Centrality] $\textrm{min}\left\{\eta_{j_1},\eta_{j_2}\right\}<\eta_{\tau_{1,2}}<\textrm{max}\left\{\eta_{j_1},\eta_{j_2}\right\}$,
\item[Higgs Mass] $115\textrm{ GeV}<m_{\tau\tau}<150\textrm{ GeV}$,
\item[Missing Energy] $\slashed{E}_{\textrm{proj}}-p_{\textrm{T},\nu_1+\nu_2}^{\textrm{fit}}>-6\textrm{ GeV}$,
\end{description}
where $j_1$ and $j_2$ are the leading and subleading jets, $m_{\tau\tau}$ is the unconstrained mass mentioned above, $\slashed{E}_{\textrm{proj}}$ is the projection of $\slashed{E}_{\textrm{T}}$ onto the transverse direction of the vectorial sum of two neutrinos' fitted momenta, $p_{\textrm{T},\nu_1+\nu_2}^{\textrm{fit}}$. This variable is useful because for the $Z\to\tau\tau$ events, the fitted neutrino momenta are stretched to comply to the Higgs mass constraint, resulting in a larger $p_{\textrm{T},\nu_1+\nu_2}^{\textrm{fit}}$ than $\slashed{E}_{\textrm{proj}}$.
\begin{table}[!bt]
\centering
\caption{The expected event yields for signal (in total and also in each decay mode) and background processes left after all selection cuts at the LHC with 300 fb$^{-1}$ luminosity. The QCD background yield is simply assumed to be similar to the QCD $Z\tau\tau$ process. }
\label{tab:events}
\begin{tabular}{ccccc}
\hline
Process & Signal & $Z\to\tau\tau$ & $Z\to\tau\tau$(EW) & QCD \\
\hline
Events & 131.2 & 96.4 & 9.4 & 96.4 \\ \hline
$\rho_{\phantom{1}}+\rho_{\phantom{1}}$ & 45.5 & 30.0 & 3.0 & 30.0 \\
$a_1+\rho_{\phantom{1}}$ & 30.3 & 25.8 & 1.5 & 25.8 \\
$\pi_{\phantom{1}}+\rho_{\phantom{1}}$ & 33.6 & 24.7 & 2.7 & 24.7 \\
$a_1+\pi_{\phantom{1}}$ & 11.6 & 8.2 & 1.5 & 8.2 \\
$\pi_{\phantom{1}}+\pi_{\phantom{1}}$ & 5.4 & 5.2 & 0.5 & 5.2 \\
$a_1+a_1$ & 4.8 & 2.5 & 0.2 & 2.5 \\
\hline
\end{tabular}
\end{table}

Based on the above reconstruction and selection, the expected event yields at 300 fb$^{-1}$ LHC are listed in Tab.~\ref{tab:events} for signal and background processes, from which one finds that the most important modes are those involving the $\rho$ meson. 
It should be noted that a Multi-Variate-Analysis of the search may give better signal sensitivity than the cuts proposed here, but it is beyond the scope of the current work.

\section{Matrix Element based analysis and results}

With fully reconstructed momentum for all final states, the calculation of the matrix element event by event is possible, which according to our parameterization (Eq.~\ref{equ:Lhtt}) has the following form:
\begin{equation}
\label{equ:ME}
|\mathcal{M}|^2 \propto A + B\cos2\phi + C\sin2\phi 
\end{equation}
where $A,\ B$ and $C$ are calculated based on Eq.~\ref{equ:Lhtt} and the effective Lagrangians and form factors for the $\tau$ decay vertices detailed in~\cite{Hagiwara:2012vz}, which depends on the momenta of all final state particles (up to a common normalization factor):
\begin{eqnarray}
A &=& 2(k_-\cdot p_{-})(k_+\cdot p_{-})-p_{-}^2(k_-\cdot k_+)+(p_{-}\leftrightarrow p_{+}) \nonumber \\
B &=& 2(g^{\mu\rho}g^{\nu\sigma}+g^{\mu\sigma}g^{\nu\rho}-g^{\mu\nu}g^{\rho\sigma})k_-^\mu k_+^\nu p_-^\rho p_+^\sigma \nonumber \\
C &=& 2\epsilon_{\mu\nu\rho\sigma}k_-^\mu k_+^\nu p_-^\rho p_+^\sigma
\end{eqnarray}
where $p_{\pm}$ are the momenta of $\tau^{\pm}$ and $k_{\pm}$ are defined as $k_{\pm}^\mu\equiv 2(J_{\pm}\cdot p_{\nu^\pm})J_{\pm}^\mu-J_{\pm}^2p_{\nu^\pm}$, where $J_{\pm}^\mu$ are the currents coupled to the $\tau-\nu_\tau$ fermion line:
\begin{eqnarray}
J_{\pm}^\mu(\tau^\pm\to\pi^\pm\nu) &=& p_{\pi^\pm}^\mu \nonumber \\
J_{\pm}^\mu(\tau^\pm\to\pi^\pm\pi^0\nu) &=& p_{\pi^\pm}^\mu - p_{\pi^0}^\mu \\
J_{\pm}^\mu(\tau^\pm\to\pi_1^\pm\pi_2^\pm\pi_3^\mp\nu) &=& F^{13}(q_1^\mu-q_3^\mu-G^{13}Q^\mu)+(1\leftrightarrow2) \nonumber 
\end{eqnarray}
where $Q^\mu = q_1^\mu + q_2^\mu + q_3^\mu$, $G^{i3}=\frac{Q\cdot(q_i-q_3)}{Q^2}$ and $F^{i3}$ are the form factors for $a_1$ channel~\cite{Hagiwara:2012vz}.

From the coefficients $B$ and $C$, an observable ($-\pi<\phi_{\textrm{ME}}<\pi$) can be constructed to retrieve the CP information ($\phi$):
%
%
\begin{equation}
\label{equ:phiME}
\cos(\phi_{\textrm{ME}}) = \frac{B}{\sqrt{B^2+C^2}}, ~\sin(\phi_{\textrm{ME}}) = \frac{C}{\sqrt{B^2+C^2}} 
\end{equation}
With this definition, the matrix element square has the form of:
\begin{equation}
|\mathcal{M}|^2\propto A + \sqrt{B^2+C^2}\cos(\phi_{\textrm{ME}}-2\phi)
\end{equation}

\begin{figure}[!tb]
\centering
\includegraphics[width=0.49\textwidth]{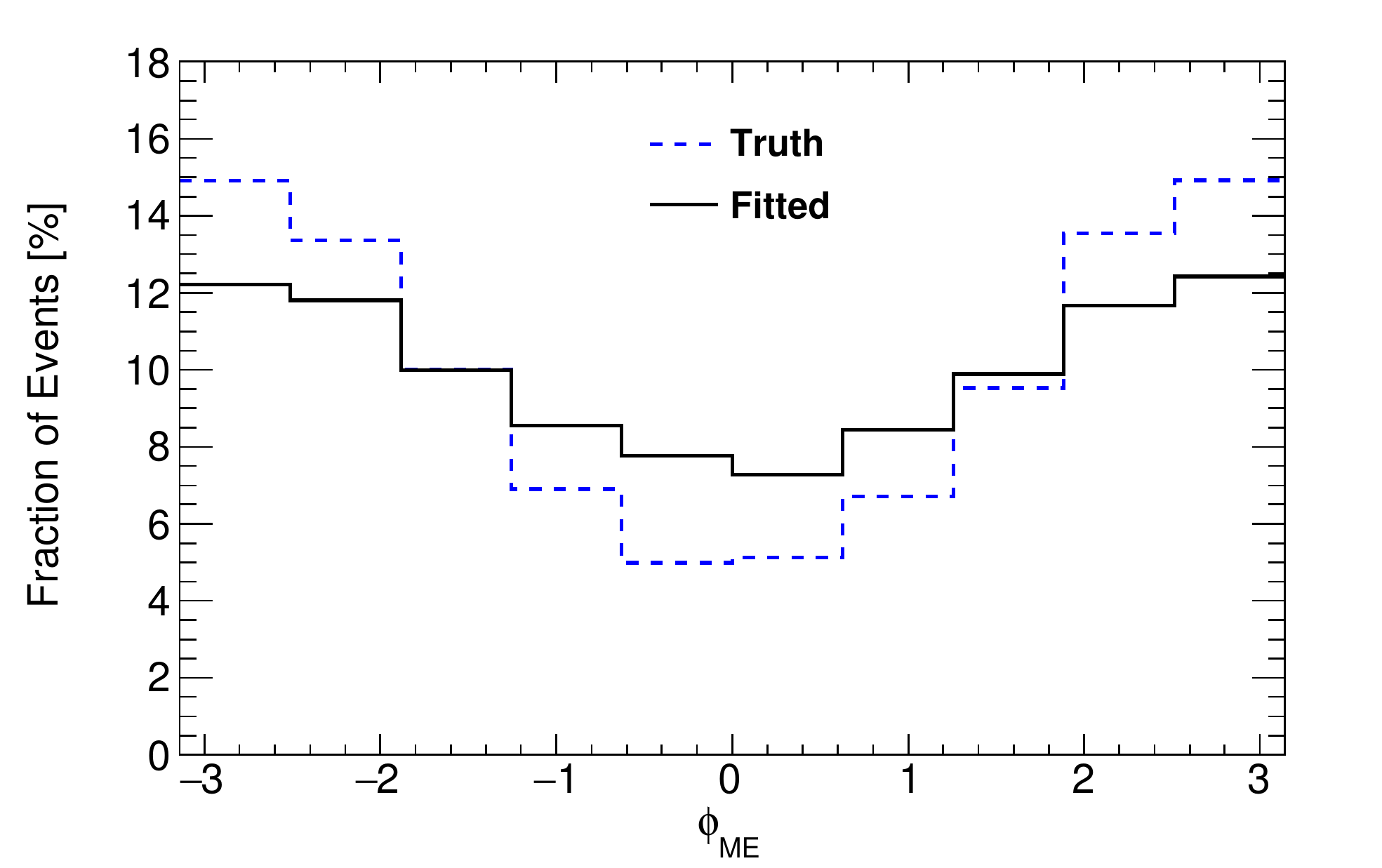}
\caption{The distribution of the angle $\phi_{\textrm{ME}}$ in truth and after the fit for the $a_1 \nu +\pi\nu$ channel in the pure CP even $h\to\tau\tau$ signal.}
\label{fig:phiMET_a1pi}
\end{figure}

Fig.~\ref{fig:phiMET_a1pi} shows the true and fitted distribution of the angle $\phi_{\textrm{ME}}$ for the $a_1 +\pi$ channel in the pure CP even case. Although the distribution is diluted after the fitting when compared to the truth, the discriminating power is still largely retained. 

\begin{figure}[!tb]
\centering
\includegraphics[width=0.50\textwidth]{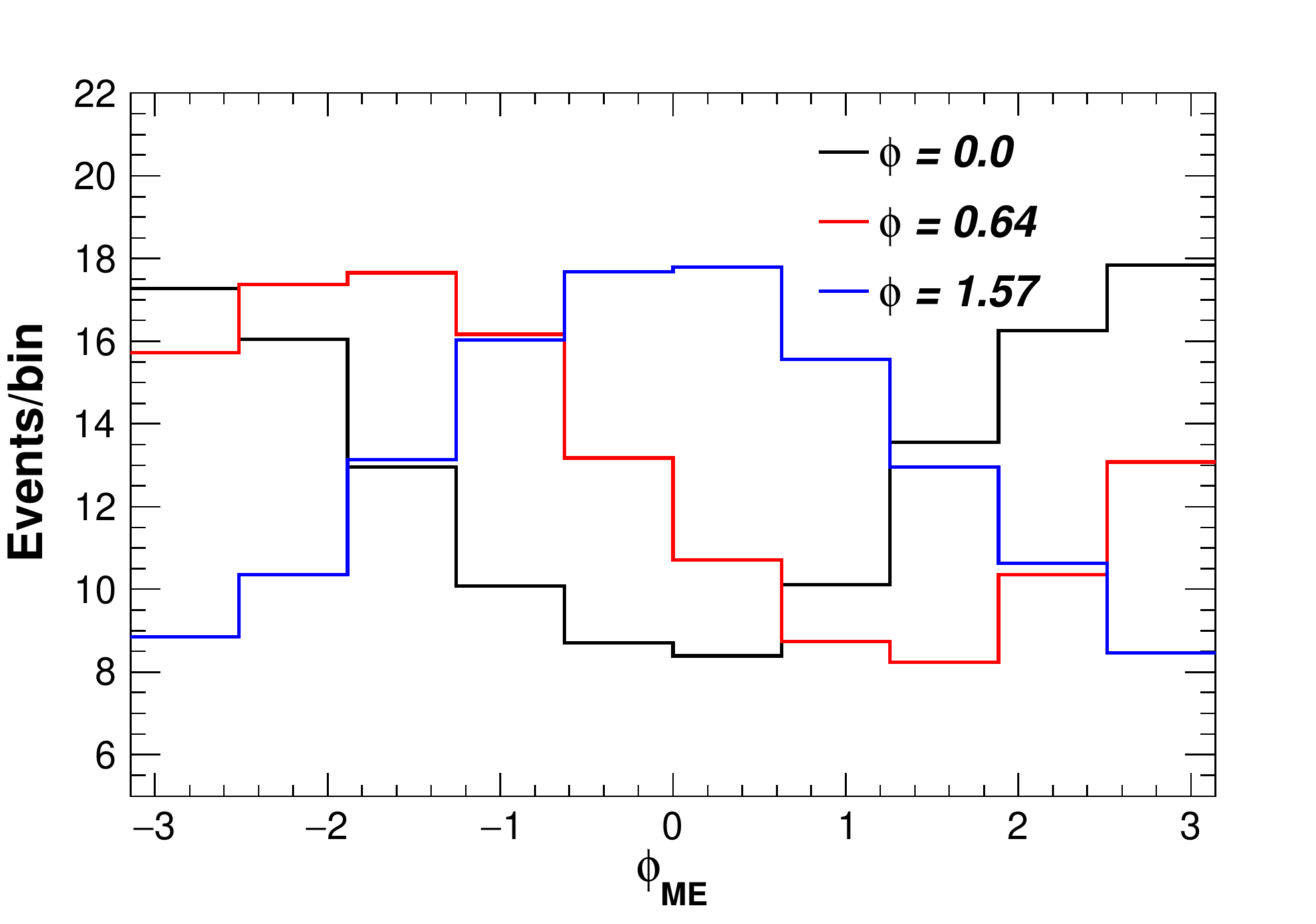}
\caption{The expected distribution of $\phi_{\textrm{ME}}$ for signal process only at the 13 TeV 300 fb$^{-1}$ LHC after all selection cuts for three different choices of CP mixing angle $\phi$: 0.0 (black line), 0.64 (red line) and 1.57 (blue line). }
\label{fig:phiME}
\end{figure}

\begin{figure}[htb]
\centering
\includegraphics[width=0.45\textwidth]{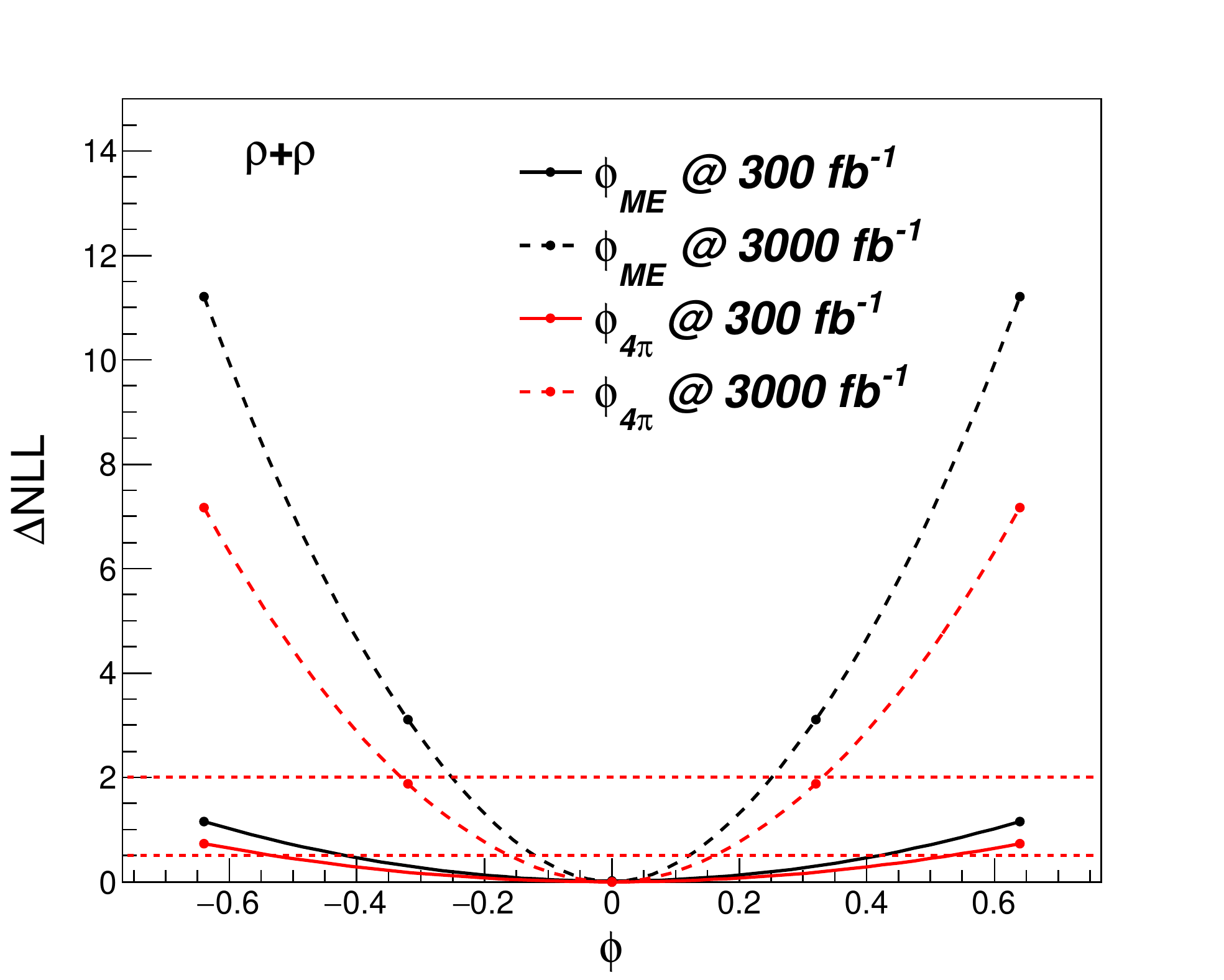} 
\put(-135, 50){\textbf{(a)}} \\
\includegraphics[width=0.45\textwidth]{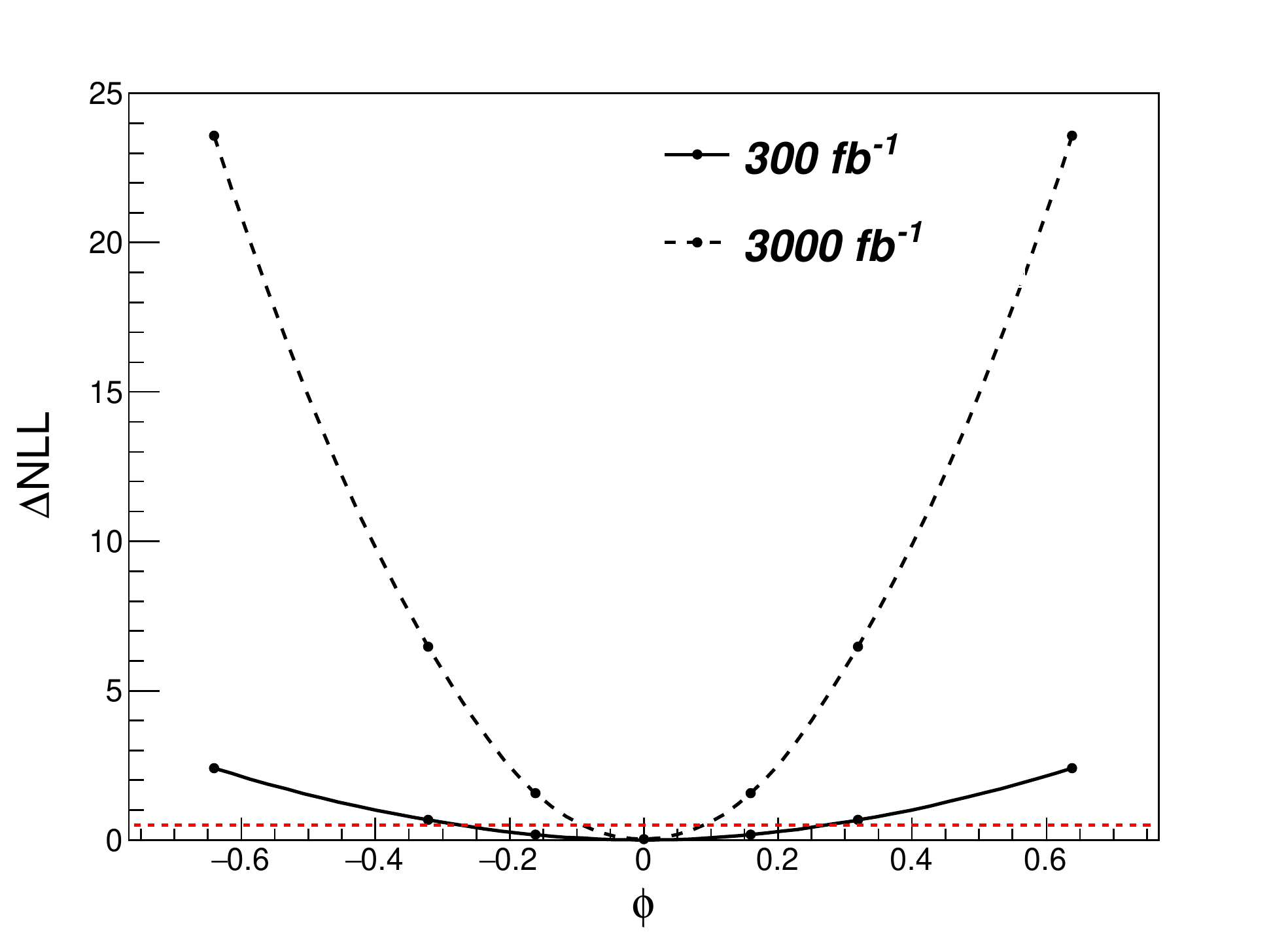}
\put(-135, 50){\textbf{(b)}}
\caption{The $\Delta$NLL as a function of the CP mixing angle $\phi$ for 300 fb$^{-1}$(solid line) and 3 ab$^{-1}$(dashed line). In Panel (a) only $\rho+\rho$ mode is used and the NLL is calculated from the expected zero-CP events in $\phi_{\textrm{ME}}$ distributions (black line), and also in $\phi_{4\pi}$ distributions (red line) from reconstructed information. 
The two horizontal dashed lines in red indicate the $1\sigma$ and $2\sigma$ confidence intervals for the $\phi$ measurement. In Panel (b), all tau decay channels are used and the NLL is calculated from the expected zero-CP events in $\phi_{\textrm{ME}}$ distributions calculated from reconstructed information. The horizontal dashed lines in red indicates the $1\sigma$ confidence interval for the $\phi$ measurement.}
\label{fig:NLL}
\end{figure}

The distribution of $\phi_{\textrm{ME}}$ for signal will be shifted according to different value of CP mixing angle $\phi$, which can be seen from Fig.~\ref{fig:phiME}, and the backgrounds have flat distributions (subject to statistical fluctuation). Note that since we have used all final states information (especially the neutrino momentum reconstructed in our method) incorporated into the matrix element to reconstruct an ``angle'' and retrieve the CP mixing information, compared with usual construction methods (one example is that used in~\cite{Han:2016bvf} and will be named as $\phi_{4\pi}$), this can achieve higher sensitivity. 
After folding in the detector efficiency and resolution effects, and imposing the selection cuts described previously, the comparison can be seen from Fig.~\ref{fig:NLL}(a) for $\rho+\rho$ mode using $\phi_{4\pi}$ as an example. 
The $y$-axis represents the difference in the Negative-Log-Likelihoods (NLL) calculated from the pseudo-data with $\phi=0$, and theoretical predictions with various values of $\phi$. The minimum NLL with $\phi=0$ hypothesis is subtracted, and the horizontal $\Delta$NLL=0.5 (2.0) line indicates the $1\sigma$ ($2\sigma$) confidence interval on the $\phi$ measurement (between intersection points with the $\Delta$NLL curves).

The result combining all decay channels listed above is presented in Fig.~\ref{fig:NLL}(b) for 300 fb$^{-1}$ (solid line) and also 3 ab$^{-1}$ (dashed line) luminosity. The 1-$\sigma$ precision at 300 fb$^{-1}$ can reach 15.5$^\circ$ (0.27 rad.) and can further be pushed down to 5.2$^\circ$ (0.09 rad.) at 3 ab$^{-1}$.

\begin{figure}[!hbt]
\centering
\includegraphics[width=0.35\textwidth]{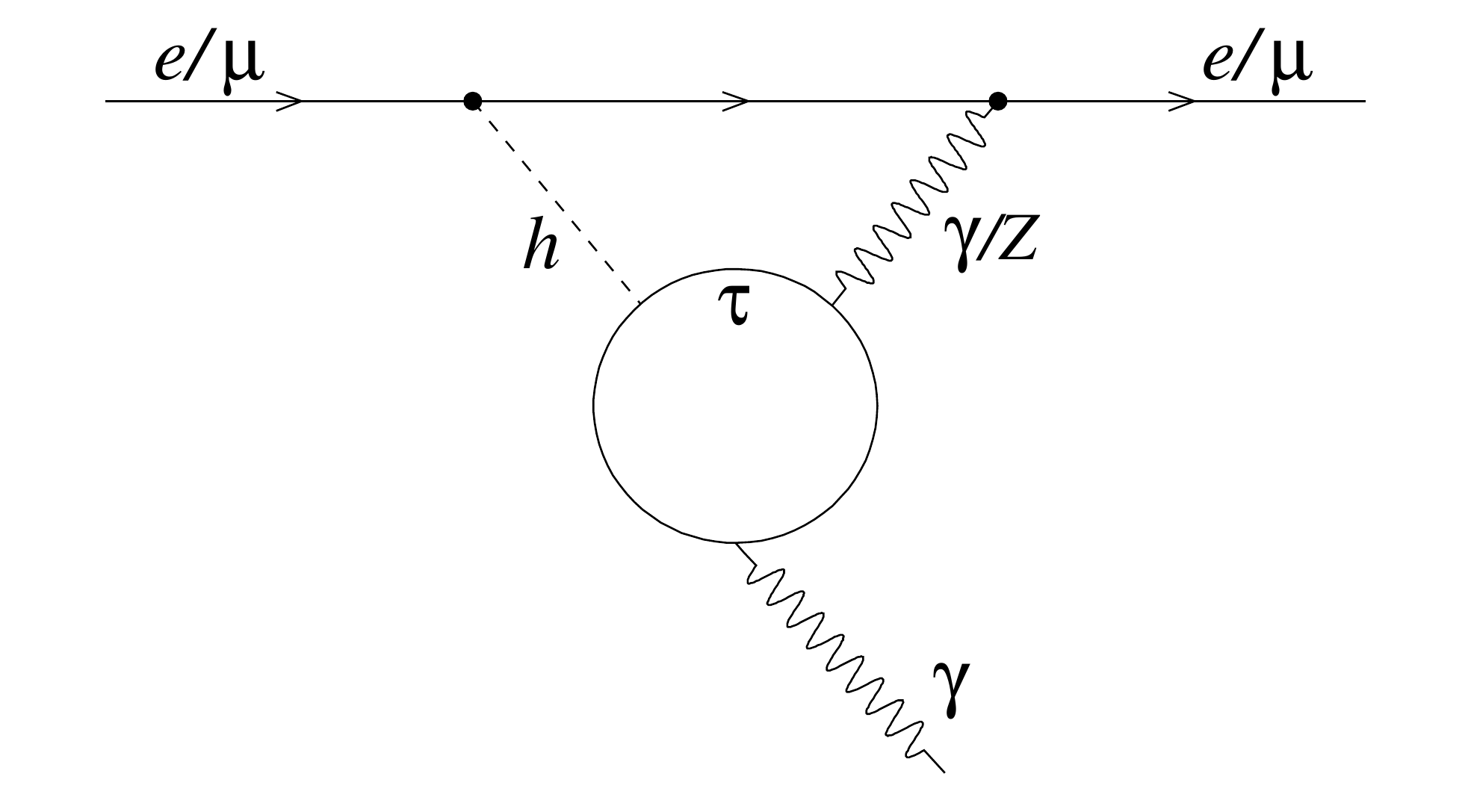}\\
\includegraphics[width=0.35\textwidth]{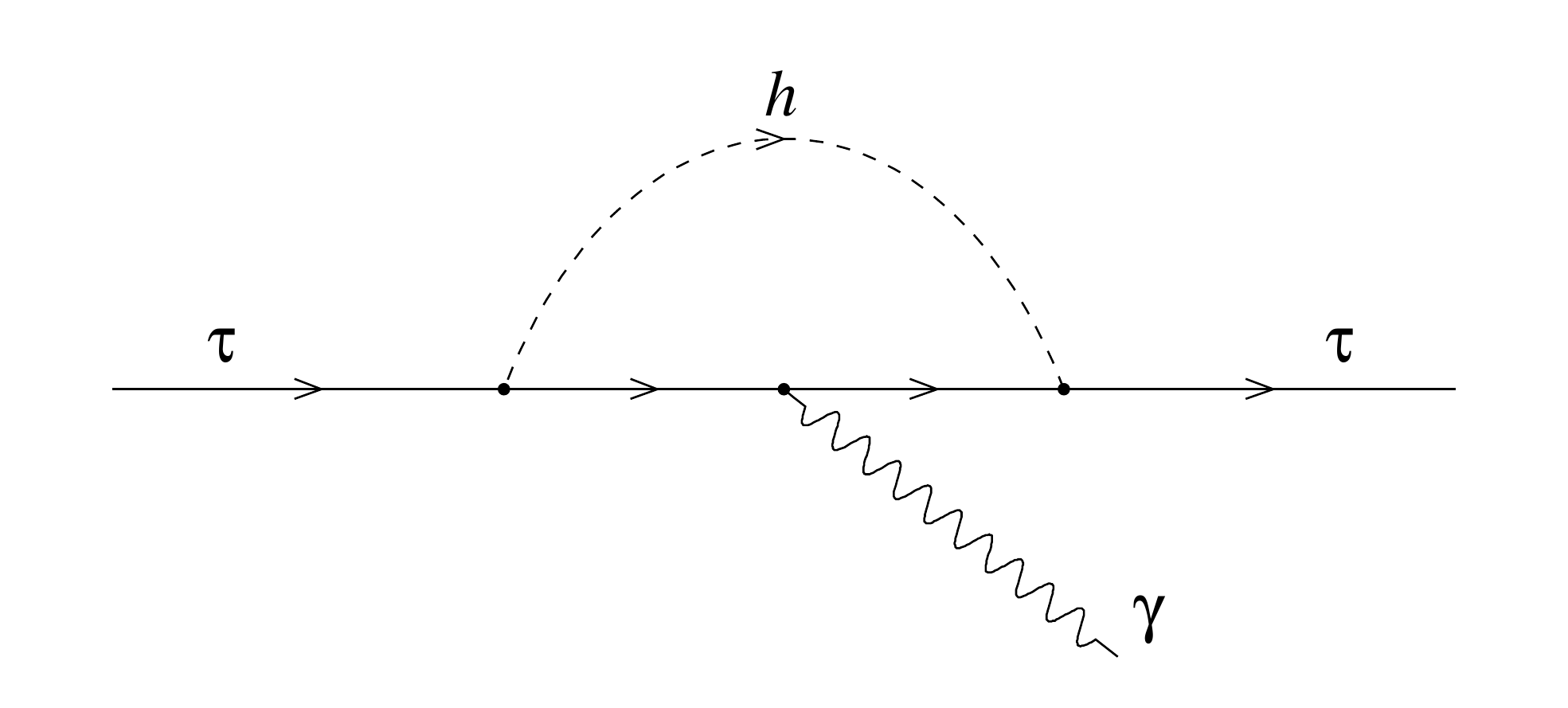}
\caption{The Feynman diagrams mediated by the 2-loop Bar-Zee process with $\tau$ lepton involved for the electron or muon EDM (top), and by the 1-loop Higgs scalar process for the tau EDM (bottom).}
\label{fig:feyn}
\end{figure}

\begin{figure}[!hbt]
\centering
\includegraphics[width=0.8\linewidth]{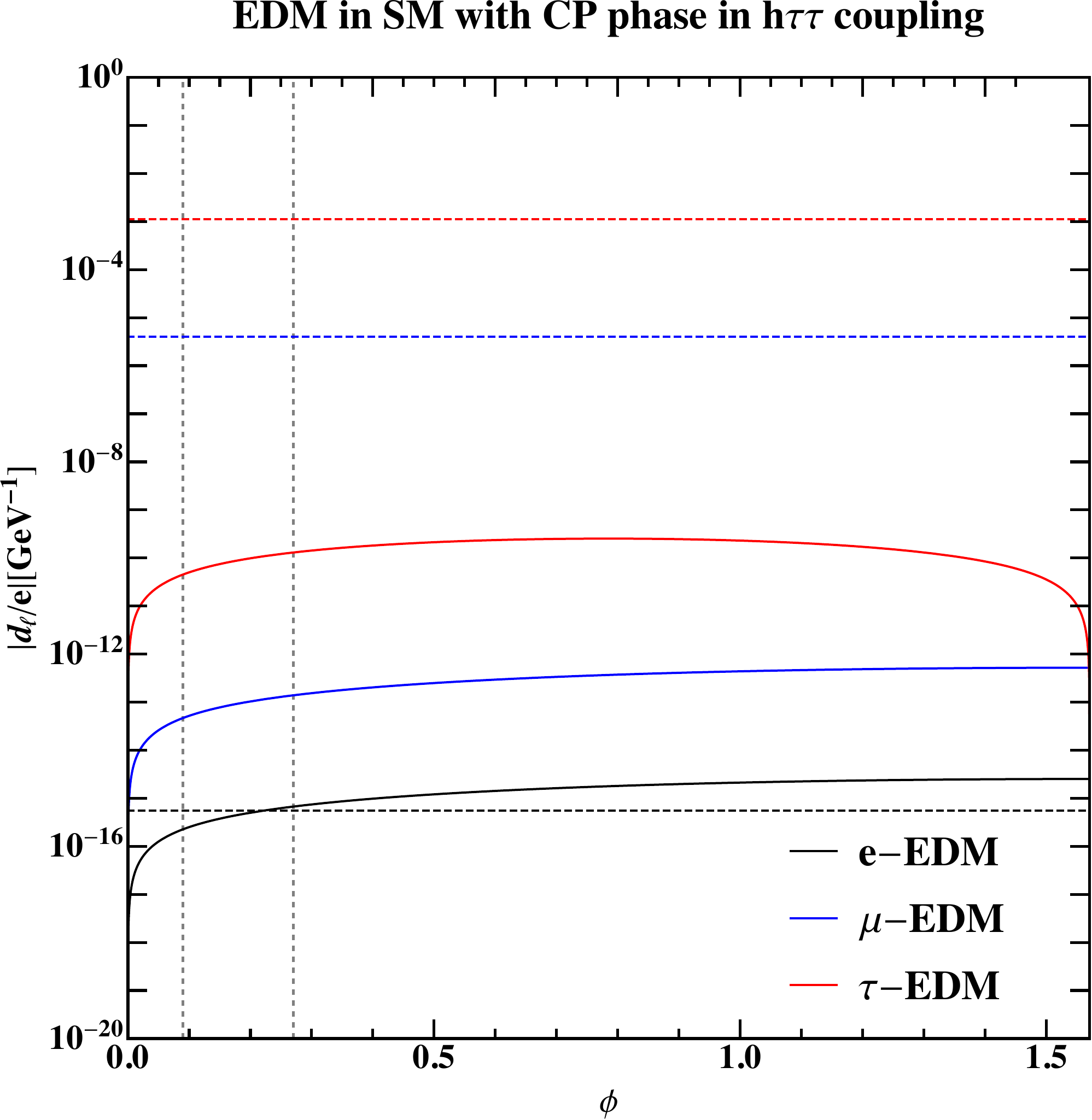}
\caption{The constraints from lepton EDM measurement. The solid lines are the corresponding EDM (black: electron, blue: muon, red: tau) induced by CP violating $h\tau\tau$ coupling. The dashed lines are the corresponding upper limits from EDM measurement. The vertical gray dashed lines are the precision from direct measurements at LHC (for 300 fb$^{-1}$ and 3  ab$^{-1}$).}
\label{fig:LepEDM}
\end{figure}

The CP-violation phase $\phi$ in the $h\tau\tau$ coupling can also induce electric dipole momentum (EDM) for the electron or muon (two-loop Bar-Zee diagrams), and tau lepton (one-loop Higgs diagram), as shown in Fig. \ref{fig:feyn}. Explicitly,
%
\begin{equation}
\label{equ:LepEDM}
d_{\tau}^{\text{1-loop}} = -\frac{e m_\tau y_\tau^2 \sin2\phi}{16\pi^2} \int_0^1 dx \frac{ x^2 }{m_\tau^2x^2+m_h^2(1-x)}, \\
\end{equation}
and for the two-loop diagrams, the formulas in \cite{BarZee_2loop} are used.
Electron and muon EDMs have already been measured to high precision, which can put stringent limit on the CP phase ($\phi$). However, if we assume that this CP phase only appears in $h\tau\tau$ coupling, the EDM measurement is less sensitive than the directly detection which is shown in Fig.~\ref{fig:LepEDM}, where the solid lines are the lepton EDM induced by CP violating $h\tau\tau$ coupling for electron (black), muon (blue) and tau (red) respectively. The dashed lines are the corresponding upper limits from EDM measurement~\cite{Andreev:2018ayy,Tanabashi:2018oca}, and the vertical gray dashed lines are the precision from direct measurement for 300 fb$^{-1}$ and 3 ab$^{-1}$ respectively. It is clear that, unless there is other CP violation physics going into the loops, the indirect detection of the CP violation effect in the tau sector from EDM measurements is not as sensitive as the direct search, even from the most precisely measured electron EDM\footnote{It is assumed that only $h\tau\tau$ coupling contributes to the electron EDM in the Bar-Zee loops. If other couplings, e.g. $htt$, also contribute, then there can be cancellation among them, such that the combined EDM effect complies with the experiment.}.

\section{Conclusion}

In conclusion, a new method is described in this paper to approximately reconstruct the neutrinos from the tau decays in the $h\tau\tau$ interaction with high precision at the LHC. The reconstructed neutrinos are used as inputs to the matrix element calculation which can be used to detect the CP-violation effect in the $h\tau\tau$ interaction. Under this framework, all major hadronic tau decay modes are included, and the VBF production region of the $H\to\tau\tau$ process is used, which has a much better signal purity than the gluon-gluon-fusion production process of Higgs. With detailed detector simulation, it is predicted that at 13 TeV LHC with 300  fb$^{-1}$ (3  ab$^{-1}$) integrated luminosity, a precision up to 15.5$^\circ$ (5.2$^\circ$) can be achieved for the CP-mixing angle ($\phi$) measurement, significantly improving previous predictions using only particular tau decay modes and/or partial event reconstructions. 
This result will provide a much more precise measurement of CP-violation effect in the $h\tau\tau$ coupling, which outperforms the sensitivity from lepton EDM searches up to date significantly.

\section*{Acknowledgments}
X. Chen wants to thank for the support from the National Thousand Young Talents program and the NSFC of China. Y. Wu is supported by the Natural Sciences and Engineering Research Council of Canada.


\bibliographystyle{elsarticle-num}
\bibliography{references}

\begin{thebibliography}{10}
\expandafter\ifx\csname url\endcsname\relax
  \def\url#1{\texttt{#1}}\fi
\expandafter\ifx\csname urlprefix\endcsname\relax\def\urlprefix{URL }\fi
\expandafter\ifx\csname href\endcsname\relax
  \def\href#1#2{#2} \def\path#1{#1}\fi

\bibitem{ATLAS-hzzCP}
G.~Aad, et~al., {Evidence for the spin-0 nature of the Higgs boson using ATLAS
  data}, Phys. Lett. B726 (2013) 120--144.
\newblock \href {http://arxiv.org/abs/1307.1432} {\path{arXiv:1307.1432}},
  \href {https://doi.org/10.1016/j.physletb.2013.08.026}
  {\path{doi:10.1016/j.physletb.2013.08.026}}.

\bibitem{CMS-hzzCP1}
S.~Chatrchyan, et~al., {Study of the Mass and Spin-Parity of the Higgs Boson
  Candidate Via Its Decays to Z Boson Pairs}, Phys. Rev. Lett. 110~(8) (2013)
  081803.
\newblock \href {http://arxiv.org/abs/1212.6639} {\path{arXiv:1212.6639}},
  \href {https://doi.org/10.1103/PhysRevLett.110.081803}
  {\path{doi:10.1103/PhysRevLett.110.081803}}.

\bibitem{CMS-hzzCP2}
S.~Chatrchyan, et~al., {Measurement of the properties of a Higgs boson in the
  four-lepton final state}, Phys. Rev. D89~(9) (2014) 092007.
\newblock \href {http://arxiv.org/abs/1312.5353} {\path{arXiv:1312.5353}},
  \href {https://doi.org/10.1103/PhysRevD.89.092007}
  {\path{doi:10.1103/PhysRevD.89.092007}}.

\bibitem{Han:2016bvf}
T.~Han, S.~Mukhopadhyay, B.~Mukhopadhyaya, Y.~Wu, {Measuring the CP property of
  Higgs coupling to tau leptons in the VBF channel at the LHC}, JHEP 05 (2017)
  128.
\newblock \href {http://arxiv.org/abs/1612.00413} {\path{arXiv:1612.00413}},
  \href {https://doi.org/10.1007/JHEP05(2017)128}
  {\path{doi:10.1007/JHEP05(2017)128}}.

\bibitem{cepc_cp}
X.~Chen, Y.~Wu, {Search for CP-Violation effects in the $h\to \tau\tau$ decay
  with future $e^+e^-$ colliders}, Eur. Phys. J. C77 (2017) 697.
\newblock \href {http://arxiv.org/abs/1703.04855} {\path{arXiv:1703.04855}}.

\bibitem{Hagiwara:2016zqz}
K.~Hagiwara, K.~Ma, S.~Mori, {Probing CP violation in $h\to \tau^{-}\tau^{+}$
  at the LHC}, Phys. Rev. Lett. 118~(17) (2017) 171802.
\newblock \href {http://arxiv.org/abs/1609.00943} {\path{arXiv:1609.00943}},
  \href {https://doi.org/10.1103/PhysRevLett.118.171802}
  {\path{doi:10.1103/PhysRevLett.118.171802}}.

\bibitem{DellAquila:1988bko}
J.~R. Dell'Aquila, C.~A. Nelson, {{CP} Determination for New Spin Zero Mesons
  by the $\bar{\tau} \tau$ Decay Mode}, Nucl. Phys. B320 (1989) 61--85.
\newblock \href {https://doi.org/10.1016/0550-3213(89)90211-3}
  {\path{doi:10.1016/0550-3213(89)90211-3}}.

\bibitem{He:1993fd}
X.-G. He, J.~P. Ma, B.~McKellar, {CP violation in Higgs decays}, Mod. Phys.
  Lett. A9 (1994) 205--210.
\newblock \href {http://arxiv.org/abs/hep-ph/9302230}
  {\path{arXiv:hep-ph/9302230}}, \href
  {https://doi.org/10.1142/S0217732394000228}
  {\path{doi:10.1142/S0217732394000228}}.

\bibitem{Hayreter:2016kyv}
A.~Hayreter, X.-G. He, G.~Valencia, {CP violation in $h\to \tau\tau$ and LFV
  $h\to \mu\tau$}, Phys. Lett. B760 (2016) 175--177.
\newblock \href {http://arxiv.org/abs/1603.06326} {\path{arXiv:1603.06326}},
  \href {https://doi.org/10.1016/j.physletb.2016.06.066}
  {\path{doi:10.1016/j.physletb.2016.06.066}}.

\bibitem{Hayreter:2016aex}
A.~Hayreter, X.-G. He, G.~Valencia, {Yukawa sector for lepton flavor violating
  in $h\to \mu\tau$ and CP violation in $h\to \tau\tau$}, Phys. Rev. D94~(7)
  (2016) 075002.
\newblock \href {http://arxiv.org/abs/1606.00951} {\path{arXiv:1606.00951}},
  \href {https://doi.org/10.1103/PhysRevD.94.075002}
  {\path{doi:10.1103/PhysRevD.94.075002}}.

\bibitem{Bower:2002zx}
G.~R. Bower, T.~Pierzchala, Z.~Was, M.~Worek, {Measuring the Higgs boson's
  parity using tau $\to$ rho nu}, Phys. Lett. B543 (2002) 227--234.
\newblock \href {http://arxiv.org/abs/hep-ph/0204292}
  {\path{arXiv:hep-ph/0204292}}, \href
  {https://doi.org/10.1016/S0370-2693(02)02445-0}
  {\path{doi:10.1016/S0370-2693(02)02445-0}}.

\bibitem{Desch:2003mw}
K.~Desch, Z.~Was, M.~Worek, {Measuring the Higgs boson parity at a linear
  collider using the tau impact parameter and tau $\to$ rho nu decay}, Eur.
  Phys. J. C29 (2003) 491--496.
\newblock \href {http://arxiv.org/abs/hep-ph/0302046}
  {\path{arXiv:hep-ph/0302046}}, \href
  {https://doi.org/10.1140/epjc/s2003-01231-4}
  {\path{doi:10.1140/epjc/s2003-01231-4}}.

\bibitem{Harnik:2013aja}
R.~Harnik, A.~Martin, T.~Okui, R.~Primulando, F.~Yu, {Measuring CP violation in
  $h \to \tau^+ \tau^-$ at colliders}, Phys. Rev. D88~(7) (2013) 076009.
\newblock \href {http://arxiv.org/abs/1308.1094} {\path{arXiv:1308.1094}},
  \href {https://doi.org/10.1103/PhysRevD.88.076009}
  {\path{doi:10.1103/PhysRevD.88.076009}}.

\bibitem{Berge:2008wi}
S.~Berge, W.~Bernreuther, J.~Ziethe, {Determining the CP parity of Higgs bosons
  at the LHC in their tau decay channels}, Phys. Rev. Lett. 100 (2008) 171605.
\newblock \href {http://arxiv.org/abs/0801.2297} {\path{arXiv:0801.2297}},
  \href {https://doi.org/10.1103/PhysRevLett.100.171605}
  {\path{doi:10.1103/PhysRevLett.100.171605}}.

\bibitem{Berge:2008dr}
S.~Berge, W.~Bernreuther, {Determining the CP parity of Higgs bosons at the LHC
  in the tau to 1-prong decay channels}, Phys. Lett. B671 (2009) 470--476.
\newblock \href {http://arxiv.org/abs/0812.1910} {\path{arXiv:0812.1910}},
  \href {https://doi.org/10.1016/j.physletb.2008.12.065}
  {\path{doi:10.1016/j.physletb.2008.12.065}}.

\bibitem{Berge:2011ij}
S.~Berge, W.~Bernreuther, B.~Niepelt, H.~Spiesberger, {How to pin down the CP
  quantum numbers of a Higgs boson in its tau decays at the LHC}, Phys. Rev.
  D84 (2011) 116003.
\newblock \href {http://arxiv.org/abs/1108.0670} {\path{arXiv:1108.0670}},
  \href {https://doi.org/10.1103/PhysRevD.84.116003}
  {\path{doi:10.1103/PhysRevD.84.116003}}.

\bibitem{Berge:2013jra}
S.~Berge, W.~Bernreuther, H.~Spiesberger, {Higgs CP properties using the $\tau$
  decay modes at the ILC}, Phys. Lett. B727 (2013) 488--495.
\newblock \href {http://arxiv.org/abs/1308.2674} {\path{arXiv:1308.2674}},
  \href {https://doi.org/10.1016/j.physletb.2013.11.006}
  {\path{doi:10.1016/j.physletb.2013.11.006}}.

\bibitem{Berge:2015nua}
S.~Berge, W.~Bernreuther, S.~Kirchner, {Prospects of constraining the Higgs
  boson’s CP nature in the tau decay channel at the LHC}, Phys. Rev. D92
  (2015) 096012.
\newblock \href {http://arxiv.org/abs/1510.03850} {\path{arXiv:1510.03850}},
  \href {https://doi.org/10.1103/PhysRevD.92.096012}
  {\path{doi:10.1103/PhysRevD.92.096012}}.

\bibitem{Dolan:2014upa}
M.~J. Dolan, P.~Harris, M.~Jankowiak, M.~Spannowsky, {Constraining
  $CP$-violating Higgs Sectors at the LHC using gluon fusion}, Phys. Rev. D90
  (2014) 073008.
\newblock \href {http://arxiv.org/abs/1406.3322} {\path{arXiv:1406.3322}},
  \href {https://doi.org/10.1103/PhysRevD.90.073008}
  {\path{doi:10.1103/PhysRevD.90.073008}}.

\bibitem{Askew:2015mda}
A.~Askew, et~al., {Prospect for measuring the CP phase in the $h\tau\tau$
  coupling at the LHC}, Phys. Rev. D91 (2015) 075014.
\newblock \href {http://arxiv.org/abs/1501.03156} {\path{arXiv:1501.03156}},
  \href {https://doi.org/10.1103/PhysRevD.91.075014}
  {\path{doi:10.1103/PhysRevD.91.075014}}.

\bibitem{Kuhn}
J.~H. Kuhn, {Tau kinematics from impact parameters}, Phys. Lett. B313 (1993)
  458.
\newblock \href {http://arxiv.org/abs/9307269} {\path{arXiv:9307269}}, \href
  {https://doi.org/10.1016/0370-2693(93)90019-E}
  {\path{doi:10.1016/0370-2693(93)90019-E}}.

\bibitem{imp1}
A.~Rouge, {CP violation in a light Higgs boson decay from tau-spin correlations
  at a linear collider}, Phys. Lett. B619 (2005) 43.
\newblock \href {http://arxiv.org/abs/0505014} {\path{arXiv:0505014}}, \href
  {https://doi.org/10.1016/j.physletb.2005.05.076}
  {\path{doi:10.1016/j.physletb.2005.05.076}}.

\bibitem{imp2}
D.~Jeans, {Tau lepton reconstruction at collider experiments using impact
  parameters}, Nucl. Instrum. Meth. A810 (2016) 51.
\newblock \href {http://arxiv.org/abs/1507.01700} {\path{arXiv:1507.01700}},
  \href {https://doi.org/10.1016/j.nima.2015.11.030}
  {\path{doi:10.1016/j.nima.2015.11.030}}.

\bibitem{ee_to_WW}
Y.~S. Tsai, A.~C. Hearn, {The Differential Cross Section for $e^+e^-\to
  W^+W^-\to e^-+\bar{\nu}_e+\mu^+ +\nu_\mu$}, Phys. Rev. 140 (1965) B721.
\newblock \href {https://doi.org/10.1103/PhysRev.140.B721}
  {\path{doi:10.1103/PhysRev.140.B721}}.

\bibitem{CMS_htt}
A.~M. Sirunyan, et~al., {Observation of the Higgs boson decay to a pair of tau
  leptons with the CMS detector}, Phys. Lett. B779 (2018) 283.
\newblock \href {http://arxiv.org/abs/1708.00373} {\path{arXiv:1708.00373}},
  \href {https://doi.org/10.1016/j.physletb.2018.02.004}
  {\path{doi:10.1016/j.physletb.2018.02.004}}.

\bibitem{ATLAS_htt}
G.~Aad, et~al., {Cross-section measurements of the Higgs boson decaying to a
  pair of tau leptons in proton–proton collisions at $\sqrt{s}$=13 TeV with
  the ATLAS detector}, ATLAS-CONF-2018-021.

\bibitem{Englert:2015dlp}
C.~Englert, O.~Mattelaer, M.~Spannowsky, {Measuring the Higgs-bottom coupling
  in weak boson fusion}, Phys. Lett. B756 (2016) 103--108.
\newblock \href {http://arxiv.org/abs/1512.03429} {\path{arXiv:1512.03429}},
  \href {https://doi.org/10.1016/j.physletb.2016.02.074}
  {\path{doi:10.1016/j.physletb.2016.02.074}}.

\bibitem{Powheg}
P.~Nason, {A New method for combining NLO QCD with shower Monte Carlo
  algorithms}, JHEP 11 (2004) 040.
\newblock \href {http://arxiv.org/abs/hep-ph/0409146}
  {\path{arXiv:hep-ph/0409146}}, \href
  {https://doi.org/10.1088/1126-6708/2004/11/040}
  {\path{doi:10.1088/1126-6708/2004/11/040}}.

\bibitem{Ball:2014uwa}
R.~D. Ball, et~al., {Parton distributions for the LHC Run II}, JHEP 04 (2015)
  040.
\newblock \href {http://arxiv.org/abs/1410.8849} {\path{arXiv:1410.8849}},
  \href {https://doi.org/10.1007/JHEP04(2015)040}
  {\path{doi:10.1007/JHEP04(2015)040}}.

\bibitem{Pythia8}
T.~Sjöstrand, S.~Ask, J.~R. Christiansen, R.~Corke, N.~Desai, P.~Ilten,
  S.~Mrenna, S.~Prestel, C.~O. Rasmussen, P.~Z. Skands, {An Introduction to
  PYTHIA 8.2}, Comput. Phys. Commun. 191 (2015) 159--177.
\newblock \href {http://arxiv.org/abs/1410.3012} {\path{arXiv:1410.3012}},
  \href {https://doi.org/10.1016/j.cpc.2015.01.024}
  {\path{doi:10.1016/j.cpc.2015.01.024}}.

\bibitem{MG5}
J.~Alwall, R.~Frederix, S.~Frixione, V.~Hirschi, F.~Maltoni, O.~Mattelaer,
  H.~S. Shao, T.~Stelzer, P.~Torrielli, M.~Zaro, {The automated computation of
  tree-level and next-to-leading order differential cross sections, and their
  matching to parton shower simulations}, JHEP 07 (2014) 079.
\newblock \href {http://arxiv.org/abs/1405.0301} {\path{arXiv:1405.0301}},
  \href {https://doi.org/10.1007/JHEP07(2014)079}
  {\path{doi:10.1007/JHEP07(2014)079}}.

\bibitem{Ball:2013hta}
R.~D. Ball, V.~Bertone, S.~Carrazza, L.~Del~Debbio, S.~Forte, A.~Guffanti,
  N.~P. Hartland, J.~Rojo, {Parton distributions with QED corrections}, Nucl.
  Phys. B877 (2013) 290--320.
\newblock \href {http://arxiv.org/abs/1308.0598} {\path{arXiv:1308.0598}},
  \href {https://doi.org/10.1016/j.nuclphysb.2013.10.010}
  {\path{doi:10.1016/j.nuclphysb.2013.10.010}}.

\bibitem{CKKW-L}
L.~Lonnblad, {Correcting the color dipole cascade model with fixed order matrix
  elements}, JHEP 05 (2002) 046.
\newblock \href {http://arxiv.org/abs/hep-ph/0112284}
  {\path{arXiv:hep-ph/0112284}}, \href
  {https://doi.org/10.1088/1126-6708/2002/05/046}
  {\path{doi:10.1088/1126-6708/2002/05/046}}.

\bibitem{Z_xsec}
R.~Gavin, Y.~Li, F.~Petriello, S.~Quackenbush, {FEWZ 2.0: A code for hadronic Z
  production at next-to-next-to-leading order}, Comput. Phys. Commun. 182
  (2011) 2388--2403.
\newblock \href {http://arxiv.org/abs/1011.3540} {\path{arXiv:1011.3540}},
  \href {https://doi.org/10.1016/j.cpc.2011.06.008}
  {\path{doi:10.1016/j.cpc.2011.06.008}}.

\bibitem{Delphes}
J.~de~Favereau, C.~Delaere, P.~Demin, A.~Giammanco, V.~Lemaître, A.~Mertens,
  M.~Selvaggi, {DELPHES 3, A modular framework for fast simulation of a generic
  collider experiment}, JHEP 02 (2014) 057.
\newblock \href {http://arxiv.org/abs/1307.6346} {\path{arXiv:1307.6346}},
  \href {https://doi.org/10.1007/JHEP02(2014)057}
  {\path{doi:10.1007/JHEP02(2014)057}}.

\bibitem{HL_LHC}
{Apollinari G. and others},
  \href{https://cds.cern.ch/record/2284929}{{High-Luminosity Large Hadron
  Collider (HL-LHC) : Technical Design Report V. 0.1, }}\href
  {https://doi.org/10.23731/CYRM-2017-004} {\path{doi:10.23731/CYRM-2017-004}}.
\newline\urlprefix\url{https://cds.cern.ch/record/2284929}

\bibitem{tracking}
\href{http://cds.cern.ch/record/2222304}{{Expected Performance of the ATLAS
  Inner Tracker at the High-Luminosity LHC}}, Tech. Rep. ATL-PHYS-PUB-2016-025,
  CERN, Geneva (Oct 2016).
\newline\urlprefix\url{http://cds.cern.ch/record/2222304}

\bibitem{antikt}
M.~Cacciari, G.~P. Salam, G.~Soyez, {The Anti-k(t) jet clustering algorithm},
  JHEP 04 (2008) 063.
\newblock \href {http://arxiv.org/abs/0802.1189} {\path{arXiv:0802.1189}},
  \href {https://doi.org/10.1088/1126-6708/2008/04/063}
  {\path{doi:10.1088/1126-6708/2008/04/063}}.

\bibitem{TauCP_HL}
\href{https://twiki.cern.ch/twiki/bin/view/AtlasPublic/TauPublicResults#Performance_Plot_for_HL_LHC_Work}{{Performance
  Plot for HL-LHC Workshop 2017 (October 2017)}}.
\newline\urlprefix\url{https://twiki.cern.ch/twiki/bin/view/AtlasPublic/TauPublicResults#Performance_Plot_for_HL_LHC_Work}

\bibitem{tausub1}
G.~Aad, et~al., {Reconstruction of hadronic decay products of tau leptons with
  the ATLAS experiment}, Eur. Phys. J. C76~(5) (2016) 295.
\newblock \href {http://arxiv.org/abs/1512.05955} {\path{arXiv:1512.05955}},
  \href {https://doi.org/10.1140/epjc/s10052-016-4110-0}
  {\path{doi:10.1140/epjc/s10052-016-4110-0}}.

\bibitem{tausub2}
V.~Khachatryan, et~al., {Reconstruction and identification of $\tau$ lepton
  decays to hadrons and $\nu_\tau$ at CMS}, JINST 11~(01) (2016) P01019.
\newblock \href {http://arxiv.org/abs/1510.07488} {\path{arXiv:1510.07488}},
  \href {https://doi.org/10.1088/1748-0221/11/01/P01019}
  {\path{doi:10.1088/1748-0221/11/01/P01019}}.

\bibitem{MINUIT}
F.~James, M.~Roos,
  \href{http://lcgapp.cern.ch/project/cls/work-packages/mathlibs/minuit}{{Minuit:
  A System for Function Minimization and Analysis of the Parameter Errors and
  Correlations}}, Comput. Phys. Commun. 10 (1975) 343--367.
\newblock \href {https://doi.org/10.1016/0010-4655(75)90039-9}
  {\path{doi:10.1016/0010-4655(75)90039-9}}.
\newline\urlprefix\url{http://lcgapp.cern.ch/project/cls/work-packages/mathlibs/minuit}

\bibitem{VBF_pheno}
D.~Z. D.~Rainwater, K.~Hagiwara, {Searching for $H\to\tau\tau$ in weak boson
  fusion at the LHC}, Phys. Rev. D59 (1998) 014037.
\newblock \href {http://arxiv.org/abs/9808468} {\path{arXiv:9808468}}, \href
  {https://doi.org/10.1103/PhysRevD.59.014037}
  {\path{doi:10.1103/PhysRevD.59.014037}}.

\bibitem{Hagiwara:2012vz}
K.~Hagiwara, T.~Li, K.~Mawatari, J.~Nakamura, {TauDecay: a library to simulate
  polarized tau decays via FeynRules and MadGraph5}, Eur. Phys. J. C73 (2013)
  2489.
\newblock \href {http://arxiv.org/abs/1212.6247} {\path{arXiv:1212.6247}},
  \href {https://doi.org/10.1140/epjc/s10052-013-2489-4}
  {\path{doi:10.1140/epjc/s10052-013-2489-4}}.

\bibitem{BarZee_2loop}
J.~B. W.~Altmannshofer, M.~Schmaltz, {Experimental constraints on the coupling
  of the Higgs boson to electrons}, JHEP 05 (2015) 125.
\newblock \href {http://arxiv.org/abs/1503.04830} {\path{arXiv:1503.04830}},
  \href {https://doi.org/10.1007/JHEP05(2015)125}
  {\path{doi:10.1007/JHEP05(2015)125}}.

\bibitem{Andreev:2018ayy}
V.~Andreev, et~al., {Improved limit on the electric dipole moment of the
  electron}, Nature 562~(7727) (2018) 355--360.
\newblock \href {https://doi.org/10.1038/s41586-018-0599-8}
  {\path{doi:10.1038/s41586-018-0599-8}}.

\bibitem{Tanabashi:2018oca}
M.~Tanabashi, et~al., {Review of Particle Physics}, Phys. Rev. D98~(3) (2018)
  030001.
\newblock \href {https://doi.org/10.1103/PhysRevD.98.030001}
  {\path{doi:10.1103/PhysRevD.98.030001}}.

\end{thebibliography}



\end{document}